\documentstyle[12pt]{article}
\input{epsf.tex}
\def\lsim{\mathrel{\rlap {\raise.5ex\hbox{$ < $}}
{\lower.5ex\hbox{$\sim$}}}}
\def\gsim{\mathrel{\rlap {\raise.5ex\hbox{$ > $}}
{\lower.5ex\hbox{$\sim$}}}}

\def\sqr#1#2{{\vcenter{\vbox{\hrule height.#2pt
        \hbox{\vrule width.#2pt height#1pt \kern#1pt
           \vrule width.#2pt}
        \hrule height.#2pt}}}}

\def\lsim{{\displaystyle
{{\raise-8pt\hbox{$ <$}}
\atop{\raise5pt\hbox{$\sim$}}}}}
\def\gsim{{\displaystyle
{{\raise-8pt\hbox{$ >$}}
\atop{\raise5pt\hbox{$\sim$}}}}}
\def\slsim{{\displaystyle
{{\raise-8pt\hbox{$\scriptstyle <$}}
\atop{\raise5pt\hbox{$\scriptstyle \sim$}}}}}
\def\sgsim{{\displaystyle
{{\raise-8pt\hbox{$\scriptstyle  >$}}
\atop{\raise5pt\hbox{$\scriptstyle \sim$}}}}}
\def\Tr{\,{\rm Tr}\, }
\def\Im{\,{\rm Im}\, }

\def\bJ{\overline{J}}
\def\bj{\overline{j}}
\def\bI{\overline{I}}
\def\bP{\overline{P}}
\def\bQ{\overline{Q}}

\def\bE{\overline{E}}
\def\bF{\overline{F}}
\def\bT{\overline{T}}
\def\bU{\overline{U}}
\def\bOmega{\overline{\Omega}}

\def\s{\sigma}
\def\a{\alpha}
\def\g{\gamma}

\def\be{\begin{equation}}
\def\ee{\end{equation}}
\def\ba{\begin{eqnarray}}
\def\ea{\end{eqnarray}}
\def\bs{\begin{subequations}}
\def\es{\end{subequations}}

\def\tb{\bar{\theta}}

\def\l{\lambda}
\def\p{\partial}

\def\rr{{\rm regular}}

\def\t{\tau}
\def\tb{\bar\tau}

\def\RR{{\rm I\!R}}
\def\zr{{Z}_{R}}
\def\zf{{Z}_{\Phi}}
\def\zh{{Z}_{H}}
\def\zc{{Z}_{C}}
\def\gs{g_{\rm string}}

\def\ap{\alpha'}
\def\za{{Z}_{F}}
\def\n{\nabla}
\def\f{{\cal F}}
\def\rr{{\cal R}}
\def\t{\tau}
\def\im{\, {\rm Im}\, \tau}
\def\b{\hat\beta}
\def\npi#1#2#3{{\bf{B#1}} (#2) #3}
\def\np#1#2#3{Nucl. Phys. {\bf{B#1}} (#2) #3}
\def\pl#1#2#3{Phys. Lett. {\bf{#1B}} (#2) #3}
\def\prl#1#2#3{Phys. Rev. Lett. {\bf{#1}} (#2) #3}

\def\nl{\hfil\break}
\def\thebibliography#1{%
\vskip 0.5cm \centerline{\bf References}
\list{%
[\arabic{enumi}]}{\settowidth\labelwidth{[#1]}
\leftmargin\labelwidth
\advance\leftmargin\labelsep
\usecounter{enumi}}
\def\newblock{\hskip .11em plus .33em minus .07em}
\sloppy\clubpenalty4000\widowpenalty4000
\sfcode`\.=1000\relax}

\begin{document}\begin{titlepage}
\begin{flushright}
CERN-TH/96-91 \\
LPTENS/96/25 \\
SISSA/60/96/EP \\
hep-th/9605011\\
\end{flushright}
\begin{centering}
\vspace{.2in}
{\bf ON THE HETEROTIC EFFECTIVE ACTION AT ONE LOOP\\
GAUGE COUPLINGS AND THE GRAVITATIONAL SECTOR\\}
\vspace{1 cm}
{E. KIRITSIS$^{\ 1}$, C. KOUNNAS$^{\ 1,\, \ast}$,
P.M. PETROPOULOS$^{\ 1,\, \ast}$ \\
\medskip
and \\
\medskip
J. RIZOS$^{\ 1,\, 2,\, \diamond}$}\\
\vskip 1cm
{$^1 $\it Theory Division, CERN}\\
{\it 1211 Geneva 23, Switzerland}\\
\medskip
{\it and}\\
\medskip
{$^2 $\it International School for Advanced Studies, SISSA}\\
{\it Via Beirut 2-4, 34013 Trieste, Italy}\\
\vspace{1.0cm}
{\bf Abstract}\\
\end{centering}
\vspace{.1in}
We present in detail the procedure for calculating the heterotic
one-loop effective action. We focus on gravitational and gauge
couplings.
We show  that the two-derivative couplings of the gravitational
sector are not renormalized at one loop when the ground state is
supersymmetric.
Arguments are presented that this non-renormalization theorem
persists to all orders in perturbation theory.
We also derive the full one-loop correction to the gauge coupling.
For a class of $N=2$ ground states, namely those that are obtained
by toroidal compactification to four dimensions of generic
six-dimensional $N=1$ models, we give an explicit formula for the
gauge-group independent thresholds, and show that these are equal
within the whole
family.
\begin{flushleft}
\medskip
{To appear in the proceedings of the {\sl 5th Hellenic
School and Workshops on Elementary Particle Physics,}
Corfu, Greece, 3--24 September 1995.}\\
\medskip
\medskip
CERN-TH/96-91 \\
LPTENS/96/25 \\
SISSA/60/96/EP \\
April 1996 \\
\end{flushleft}
\hrule width 6.7cm \vskip.1mm{\small \small \small
$^\ast$\ On leave from {\it Centre National de
la Recherche Scientifique,} France.\\
$^\diamond$\ On leave from {\it Division of Theoretical Physics,
Physics Department,}\\
$^{\ }$\ {\it University of Ioannina,} Greece.}
\end{titlepage}
\newpage
{\bf 1. Introduction}

In the past several years, there has been significant progress in
trying to compare
low energy predictions of string theory with data
\cite{minahan}--\cite{DF}.
String theory gives us the possibility of unifying gauge, Yukawa and
gravitational interactions.
The presence of supersymmetry is usually required in order to deal
with hierarchy problems (although in the context of supergravity and
strings this is not automatic, due to the presence of gravity
\cite{fkz}).
The standard folklore demands $N=1$ supersymmetry in order for the
theory to possess chiral fermions. There seem however to be flaws in
this popular wisdom \cite{kka}.

The quantities that are most easily comparable to experimental data
are effective gauge couplings of the observable sector, as well as
Yukawa couplings.
It is well known that the low-energy world is not supersymmetric.
Thus supersymmetry has to be broken spontaneously at some scale of
the order of 1 TeV (for hierarchy reasons).
Although there are ways to break supersymmetry in string theory
\cite{gau,ss,bach}, it is fair to say that none so far has yielded a
phenomenologically acceptable model.
Although the issue of supersymmetry breaking is an open problem, if
we assume
that its scale is of the order of 1 TeV and the superpartner masses
are around that scale,
then non-supersymmetric thresholds are not very important for
dimensionless couplings
(which include gauge and Yukawa couplings). Thus, it makes sense to
compute
them and compare them with data in the context of unbroken
supersymmetry.

There are several procedures to compute the one-loop corrections to
dimensionless couplings in string theory. The most powerful and
unambiguous one
was described in \cite{kkb,kkc}. It amounts to turning on
gravitational background fields that provide the ground state in
question with a mass gap $\Delta m^2$,
and further background fields (magnetic fields, curvature and
auxiliary $F$ fields) in order to
perform a background-field calculation of the relevant one-loop
corrections.

The above procedure involves the following steps:
\nl
(\romannumeral1)
We first regulate the infra-red by introducing a mass gap in the
relevant ground state. This is done by replacing the flat
four-dimensional conformal field theory with
the wormhole one, $\RR_Q \times SO(3)_{k\over 2}$ \cite{worm}.
The mass gap is given by $\Delta m^2={M_{s}^2 \over 2 (k+2)}$, where
$M_{s}={1\over \sqrt{\alpha'}}$ and $k$ is a (dimensionless)
non-negative even integer.
\nl
(\romannumeral2)
We then turn on appropriate background fields, which are exact
solutions of the string equations of motion. Such backgrounds include
curvature, magnetic fields and auxiliary $F$ fields\footnote{These
are relevant for the study of the K\"ahler potential renormalization.
For more details see \cite{pet}.}.
\nl
(\romannumeral3)
{We calculate the one-loop vacuum amplitude as a function of these
background fields.
\nl
(\romannumeral4)
We identify these background fields as solutions of the tree-level
effective action. By substituting them into the one-loop effective
action and comparing with the string calculation of the free energy,
we can extract the renormalization constants at one loop.

In the following we will apply the aforementioned techniques to the
calculation of string loop corrections for gauge and gravitational
couplings. For heterotic ground states with at least $N=1$
supersymmetry, we will demonstrate that Newton's constant is not
renormalized, and derive the full one-loop gauge coupling. We
will in particular obtain an explicit formula for the universal
part of the threshold corrections. Finally, we will show how,
for the whole class  of $N=2$ ground states
that come from two-torus compactification of six-dimensional $N=1$
theories, these thresholds are equal and fully determined as a
consequence of an anomaly-cancellation constraint in six dimensions.
This summarizes our main results.
\vskip 0.3cm
{\bf 2. Infra-red regularization and background fields}

As mentioned in the introduction, an infra-red regulated version of a
given heterotic ground state is provided by substituting
four-dimensional flat space with the $\RR_{Q}\times SO(3)_{k\over 2}$
conformal field
theory\footnote{The group $SO(3)$ is required instead of $SU(2)$ for
spin-statistics consistency \cite{kkc}.}. More details can be found
in \cite{kkb,kkc}. There is a linear dilaton in the time direction
\be
\Phi={t M_s\over \sqrt{k+2}}\, ,
\ee
necessary for making the total central charge equal to that of flat
space.
The mass gap can be read off from the left worldsheet Hamiltonian
(in the Euclidean)
\be
L_{0}=-{1\over 2}+{1\over 4(k+2)}+{p_t^2\over 2}+{j(j+1)\over k+2}+
\cdots\, ,
\ee
to be $\Delta m^2={\mu^2\over 2}$ with $\mu= {M_{s}\over
\sqrt{k+2}}$.

In this geometry there are several marginal deformations, which turn
on background fields.
For magnetic fields we use
\be
V^{\rm magn}_i\propto \left(J^3 + i : \psi^1 \psi^2 :\right)
\overline{J}_i^{\vphantom m}\, .
\ee
This turns on a magnetic field in the third space direction;
$\overline{J}_i$ is a right-moving affine current in the Cartan of
the $i$th gauge group simple factor
(picking out a single Cartan direction will be enough for our
purposes), and $J^3$ belongs to the $SO(3)_{k\over 2}$ affine Lie
algebra.
There is also a gravitational perturbation generated by
\be
V^{\rm grav}\propto \left(J^3+i:\psi^1\psi^2:\right)\bJ^3\,
{}.
\ee
The currents $J^3$, $\overline{J}^3$ and $\overline{J}_i$ are
normalized so that\footnote{Notice that there is a factor of 2
difference between the normalization used here and that used in
\cite{kkb,kkc}. Our present normalization is the one widely used in
the literature; it corresponds to the situation where the highest
root of the algebra has length
squared $\psi ^2=2$.}
\be
J^3(z)\, J^3(0)= {k\over 2z^2}+\cdots\, ,\ \ \bJ^3(\bar z)\,
\bJ^3(0)=
{k\over 2\bar z^2}+\cdots\, ,\ \ \bJ_i(\bar z)\, \bJ_i(0)= {k_i\over
\bar z^2}
+\cdots \, .
\ee
All the above perturbations are products of left times right Abelian
currents
and thus preserve conformal invariance. This implies that the new
backgrounds
satisfy the string equations of motion at tree level to all orders in
the $\ap$ expansion.

The vacuum amplitude at one loop, i.e. the free energy, in the
presence of these backgrounds can be readily calculated:
\be
\alpha'^2 F^{\rm \ string}_{\rm one \ loop}=
{1\over 2(2\pi)^4}\int_{\f}{d^2\tau\over (\im)^2} \,
D^{\rm \ string}_{\rm one \ loop}=
{1\over 2(2\pi)^4}\int_{\f}{d^2\tau\over (\im)^2} \,
\left\langle e^{-2\pi\im\, \delta
\left(L_{0}+ \overline{L}_{0}\right)}\right\rangle\, ,
\label{32}
\ee
with
\ba
\delta L_{0}=
\delta \overline{L}_{0}&=&{\sqrt{1+\f^2+\rr^2}-1\over 2}
\left(
{\left(Q+I^3\right)^2\over k+2}
+{1\over \rr^2+\f^2}
\left(\rr{\bI^3\over \sqrt{k}}+\f{\bP_i\over \sqrt{2k_{i}}}\right)^2
\right)\cr
&& +
{Q+I^3\over \sqrt{k+2}}
\left(\rr{\bI^3\over \sqrt{k}}+\f{\bP_i\over \sqrt{2k_{i}}}\right)
\, .
\label{33}
\ea
Here $I^3,\bI^3$ stand for the zero modes of the respective
$SO(3)_{k\over 2}$ currents,
$Q$ is the zero mode of the $i:\psi^1\psi^2:$ current and $\bP_i$ is
the zero mode of the $\bJ_i$ current.
We also assume that the gauge background does not correspond to an
anomalous $U(1)$. This case can also be treated, but is more
complicated.
Since anomalous $U(1)$'s are broken at scales comparable with the
string scale, their running is irrelevant for low-energy physics.
Expanding to second order in the background fields, we find:
\ba
D^{\rm \ string}_{\rm one \ loop}&=&\langle 1\rangle \cr
&&+
{8\pi^2(\im)^2\rr^2\over k(k+2)}
\left\langle
\left(Q+I^3\right)^2\left(\bI^3\right)^2-
{k\over 8\pi\im}
\left(\left(Q+I^3\right)^2+{k+2\over k}\left(\bI^3\right)^2\right)
\right\rangle \cr
&&+
{4\pi^2(\im)^2\f^2\over k_{i}(k+2)}
\left\langle
\left(Q+I^3\right)^2\bP_i^2-
{k_{i}\over 4\pi\im}
\left(\left(Q+I^3\right)^2+{k+2\over 2 k_i}\bP_i^2\right)
\right\rangle \cr
&&+\cdots
\, ,
\label{34}
\ea
where the dots stand for higher orders in $\f$ and $\rr$.

Here we will assume that our ground state has at least $N=1$
supersymmetry\footnote{The general formula in the absence of
supersymmetry can be found in \cite{kkb}.}.
In such ground states, terms in (\ref{34}) that do not contain the
helicity operator $Q$ vanish because of the presence of the fermionic
zero modes, and terms linear in $Q$ vanish due to rotational
invariance,
$\langle I^3\rangle =0$.
Thus for $N=1$ ground states, (\ref{34}) becomes
\be
D^{\rm \ string}_{\rm one \ loop}=
{8\pi^2(\im)^2\over k+2}\left\langle
Q^2\Bigg( {\rr^2\over k}
\left( \left(\bI^3\right)^2-{k\over 8\pi\im}\right)
+
{\f^2\over 2k_{i}}
\left(\bP_i^2-{k_i\over 4\pi\im}\right)
\Bigg)\right\rangle
+\cdots\, .
\label{35}
\ee
The generic $N=1$ four-dimensional vacuum amplitude has the form
\be
\langle 1\rangle={1\over \im\, |\eta|^4}\sum_{a,b=0,1}
{\vartheta{a\atopwithdelims[]b}\over \eta}\,
C{a\atopwithdelims[]b}\, \Gamma(k)=0\, ,
\label{34a}
\ee
where $C{a\atopwithdelims[]b}$ is the contribution of the internal
conformal field theory, and
\be
\Gamma(k)= {4 \sqrt{x}}
\left.{\partial\over \partial x}
\Big[\varrho(x)-\varrho(x/4)\Big]\right\vert_{x=k+2}
\ {\rm with}\ \
\varrho(x)=\sqrt{x}\sum_{m,n\in Z}e^{-{\pi x \over \im}|m+n\t|^2}
\ee
stands for the $SO(3)_{k\over 2}$ partition function
normalized so that
$\lim_{k\to \infty}\Gamma(k) = 1$. This extra factor
ensures the convergence of integrals such as those appearing in
(\ref{32}),
at large values of $\im$. Expression (\ref{34a}) allows us to recast
(\ref{35}) as
follows:
\ba
D^{\rm \ string}_{\rm one \ loop}=
-{4\pi i\over k+2}{\im\over |\eta|^4}
&&\!\!\!\!\!\!\!\!\!
\sum_{a,b=0,1}
\Bigg\{
{\f^2\over  k_{i}}\,
\partial_{\t}
\left({\vartheta{a\atopwithdelims[]b}\over\eta}\right)\,
\left(
\bP_i^2 - {k_{i}\over 4\pi{\im}}\right) C{a\atopwithdelims[]b}\,
\Gamma(k)
\cr
&&\!\!\!\!\!\!\!\!\!\!\!-
{\rr^2\over 6 k}\,
\partial_{\t}
\left({\vartheta{a\atopwithdelims[]b}\over\eta}\right)\,
C{a\atopwithdelims[]b}\,
\left(\widehat{E}_{2}+{2(k+2)\over i
\pi}\partial_{\tb}\right)\Gamma(k)
\Bigg\} +\cdots\, ,
\label{36}
\ea
where $\bP_i^2$ acts as ${i \over \pi}{\partial \over \partial \bar
\tau}$ on the
appropriate subfactor of the 32 right-moving-fermion
contribution, and
\be
\widehat{E}_{2}\equiv {6i\over \pi}\partial_{\tb}\log\left(\im \,
\bar\eta^2\right)
=\bE_2 - {3\over \pi \im} \, ;
\label{377}
\ee
$E_2$ is an Eisenstein holomorphic function
(see (\ref{61b})) and $\widehat{E}_2$ is modular-covariant of degree
2. Since we are interested in the large-$k$ limit, we can expand
(\ref{36}) in powers of $1/k$. In the next-to-leading order, the
above expression reads:
\ba
D^{\rm \ string}_{\rm one \ loop}=
-{4\pi i\over k}{\im \over |\eta|^4}\, \Gamma(k)
\sum_{a,b=0,1}&&\!\!\!\!\!\!\!\!\!\!\!
\Bigg\{
{\f^2\over  k_{i}}\,
\partial_{\t}
\left({\vartheta{a\atopwithdelims[]b}\over\eta}\right)\,
\left(
\bP_i^2 - {k_{i}\over 4\pi{\im}}\right) C{a\atopwithdelims[]b}\,
\left(1-{2\over k}\right)
\cr
&&\!\!\!\!\!\!\!\!\!\!\!-
{\rr^2\over 6 k}\,
\partial_{\t}
\left({\vartheta{a\atopwithdelims[]b}\over\eta}\right)\,
\widehat{E}_{2}
\,
C{a\atopwithdelims[]b}+{\cal O}\left({1\over k^2}\right)\Bigg\}
+\cdots\, .
\label{366}
\ea
This is the form of the one-loop density that we will use for our
subsequent calculations.
\vskip 0.3cm
{\bf 3. One-loop effective action and the gravitational sector}

The tree-level heterotic effective action is given
by\footnote{Expression (\ref{1}) holds in the $\sigma$-model frame,
which is the natural frame for perturbative string calculations.}
\be
S_{\rm tree}={1\over \alpha'}\int d^{4}x\,
\sqrt{G}\, e^{-2\Phi} \left( R + 4(\n\Phi)^2-
{1\over 12}{\widehat H}^2-\alpha'
\sum_{i,a}{1\over 4 g_{i}^2} \,
F^a_{i\, \mu\nu}\,
F_i^{a\, \mu\nu}
+ \cdots \right)\, ,
\label{1}
\ee
where the dots stand for two-derivative terms that include the
scalars and fermions, as well as higher-derivative terms.
The tree-level cosmological constant is set to zero since we consider
ground states with the appropriate value of the central charge.
The index $i$ labels the various simple components of the gauge
group,
while $a$ spans the corresponding adjoint representations.
The tree-level couplings $g_{i}$ are dimensionless and are given here
by $g_{i}={1\over \sqrt{k_{i}}}$, where $k_{i}$ are the integer
levels of the appropriate affine algebras responsible for the gauge
group. Note that the physical gauge couplings contain also the string
coupling
$\gs = \exp \langle \Phi \rangle$; this will be restored in the next
section.
We have also introduced
\be
\widehat H_{\mu\nu\rho}=
         H_{\mu\nu\rho}-\ap \sum_{i}
{1\over 2 g_i^2}\,C\,  S_{i\, \mu\nu\rho}\, ,
\label{2}
\ee
where
\be
H_{\mu\nu\rho}=\p_{\mu}B_{\nu\rho}+{\rm
cyclic\ permutations}
\label{3}
\ee
and
\be
C\, S_{i\, \mu\nu\rho}^{\hphantom a}=\sum_{a}
A^a_{i\, \mu}\,
F^a_{i\, \nu\rho}
-{1\over 3}\sum_{a,b,c}
f^{abc}_{i}\,
A^{a}_{i\, \mu}\,
A^{b}_{i\, \nu}\,
A^{c}_{i\, \rho}
+{\rm cyclic\ permutations.}
\label{4}\ee

We will now translate the gravitational and magnetic backgrounds
described in the previous section in terms of conformal field theory
into the language of the effective action.
We will use the Euler-angle parametrization of $SO(3)$.
In this parametrization we have three angles:
$\beta\in\left[0,\pi\sqrt{\ap k}\right]$ and $\alpha,\gamma\in
\left[0,2\pi\sqrt{\ap k}\right]$.
The three coordinates $\alpha$, $\beta$ and $\gamma$ as well as $t$
have dimensions of length, and $\b=\beta/\sqrt{\ap k}$ is
dimensionless. The fields $G_{\mu\nu}$, $B_{\mu\nu}$ and $\Phi$
are dimensionless, and the gauge fields have dimensions of mass.

It is not difficult to verify \cite{kkc} that the conformal field
theory backgrounds  of section 2 correspond to the following metric,
antisymmetric tensor and dilaton:
\ba
\ \ \ \ \
G_{tt}=1\, ,\ \  G_{\beta\beta}={1\over 4}
\, ,\ \
G_{\a\a}={1\over 4}
{\left(\l^2+1\right)^2-\left(8H^2\l^2+\left(\l^2-1\right)^2\right)
\cos^2 \b          \over
\left(\l^2+1+\big(\l^2-1\big)\cos\b\right)^2}
\, ,
\cr
G_{\g\g}={1\over 4}
{\left(\l^2+1\right)^2-8H^2\l^2-\left(\l^2-1\right)^2\cos^2\b\over
\left(\l^2+1+\big(\l^2-1\big)\cos\b\right)^2}
\, ,
\ \ \ \ \ \ \ \ \ \ \ \ \ \ \ \
\cr
G_{\a\g}={1\over 4}
{4\l^2\left(1-2H^2\right)\cos\b-\left(\l^4-1\right)\sin^2\b\over
\left(\l^2+1+\big(\l^2-1\big)\cos\b\right)^2}\, ;
\ \ \ \ \ \ \ \ \ \ \ \ \ \ \ \
\label{15}
\ea
\be
B_{\a\g}={1\over 4}{\l^2-1+\left(\l^2+1\right)\cos\b\over
\l^2+1+\big(\l^2-1\big)\cos\b}\, ;
\label{16}
\ee
\be
\Phi={t\over \sqrt{\ap k}}-{1\over 2}\log\Bigg(\l+{1\over
\l}+\left(\l-{1\over \l}
\right)\cos\b\Bigg)+\log\gs\, .
\label{19}
\ee
The non-vanishing gauge field components are those corresponding to
the Cartan direction that we have chosen in the $i$th simple group
factor:
\be
A^a_{i\,\a}={2\, g_i\over \sqrt{\ap}}{H\l\cos\b\over
\l^2+1+\big(\l^2-1\big)\cos\b}
\, , \ \
A^a_{i\,\g}={2\, g_i\over \sqrt{\ap}}{H\l\over
\l^2+1+\big(\l^2-1\big)\cos\b}
\, .
\label{18}
\ee
We can check that these  background fields satisfy the equations of
motion stemming from the tree-level effective action~(\ref{1}). Note
also that the exact solution (to all  orders) for the dilaton is
obtained by shifting $k\to k+2$ in eq. (\ref{19}). The relation of
the effective parameters $H$ and $\lambda$ to the conformal field
theory parameters $\f$ and $\rr$ is summarized in the following
equations:
\be
H^2={1\over 2}{\f^2\over \f^2+2\left(1+\sqrt{1+\f^2+{\cal
R}^2}\right)}
={\f^2\over
8}\Big(1+{\cal O}\left(\f^2,{\cal R}^2\right)\Big)\, ,
\label{23}
\ee
\be
\l^2= {1+\sqrt{1+\f^2+{\cal R}^2}+{\cal R}\over
1+\sqrt{1+\f^2+{\cal R}^2}-{\cal R}}
=1+{\cal R}+{\rr ^2 \over 2}+
{\cal O}\left(\f^3,{\cal R}^3\right)\, .
\label{24}
\ee

Let us now turn to the corrections that the above effective action
receives at one loop. The  bosonic part of the latter in the
universal and gauge sectors can be parametrized as
\ba
S_{\rm one \ loop}=
{1\over \alpha'}\int d^{4}x\, \sqrt{G}\,
\Bigg({\Lambda_{\rm one \ loop}\over \alpha'} +\zr
 \,R+4\zf\,(\n\Phi)^2
\ \ \ \ \ \ \ \ \ \ \ \ \ \ \ \ \ \ \ \ \ \ \ \ \ \ \ \ \ \ \
\cr \ \ \ \ \ \ \ \ \ \ \ \ \ \ \ \ \ \ \ \ \ \
-{\zh\over 12}
\bigg(H-\ap \sum_{i}
{\zc\over 2 g_i^2}\,CS^{\vphantom a}_i\bigg)^2
-\ap
\sum_{i,a}{\za\over 4 g_{i}^2} \,
F^a_{i\, \mu\nu}\,
F_i^{a\, \mu\nu}
\Bigg)
\, .
\label{5}
\ea
All renormalization coefficients $Z_{K}$ are dimensionless.
They encode the one-loop corrections to the various effective
couplings.
There are some extra couplings associated with anomalous $U(1)$'s but
we will
not consider this case here. Since the above action is the torus
contribution,
there is no overall dilaton-dependent factor. Assuming unbroken
supersymmetry amounts to a vanishing
$\Lambda_{\rm one \ loop}$.  Moreover,
we know from previous studies \cite{stie} that there is no
Chern-Simons coupling at one loop, which in turn implies that
$\zc=0$.

What we need to do next is to compute the free energy associated with
the background fields studied above.
This can be done by evaluating the corresponding one-loop action.
Since the various backgrounds do not depend on the Killing
coordinates, the  four-dimensional measure, once normalized with
respect to its flat-space limit
$V_{\rm \, fl.\, sp.}
={\sqrt{\ap k}\over 4}\int dt\, d\a \, d\g$, becomes
\be
\int d^{4}x\, \sqrt{G}\to \sqrt{1-2H^2}\int_{0}^{\pi} d\b \,{\sin
\b\over \Big(\l+{1\over \l}+\left(
\l-{1\over \l}\right)\cos \b\Big)}
\, .
\label{22}
\ee
We can now evaluate the various terms that are relevant for the
action (\ref{5}).
After some calculation we obtain to leading order in the $1/ k$
expansion:
\be
{1\over\ap}\int {d^4 x \over V_{\rm \, fl.\, sp.}}\, \sqrt{G}\,R=
{1\over k}\, \sqrt{1-2H^2}\, \left(3+2H^2\right)\left(\l+{1\over
\l}\right)=
{1\over k}\left(6-{\f^2\over 4}+{3\rr^2\over 4}+\cdots \right)\, ,
\label{25}
\ee
\be
{1\over\ap}\int {d^4 x \over V_{\rm \, fl.\, sp.}}\,  \sqrt{G}\,4
(\n\Phi)^2= {2\over k}\, \sqrt{1-2H^2}\left(\l+{1\over \l}\right)=
{1\over k}\left(4-{\f^2\over 2}+{\rr^2\over 2}+\cdots \right)\, ,
\label{26}
\ee
\be
{1\over\ap}\int {d^4 x \over V_{\rm \, fl.\, sp.}}\, \sqrt{G}\,
{H^2\over 12}={1\over k}{1\over \sqrt{1-2H^2}}\left(\l+{1\over
\l}\right) ={1\over k}\left(2+{\f^2\over 4}+{\rr^2\over 4}+\cdots
\right)\, ,
\label{27}
\ee
\be
\int {d^4 x \over V_{\rm \, fl.\, sp.}}\, \sqrt{G}\,{F_i^2\over
4g_i^2}={4\over k}\, H^2\sqrt{1-2H^2}\left(\l+{1\over
\l}\right)={1\over k}\, \f^2+\cdots \, .
\label{28}
\ee
We have used relations (\ref{23}) and (\ref{24}) in the right-hand
side above.
It should be noted that all such lowest-order contributions are of
order $1/k$. In fact the expansion in powers of $1/k$ organizes the
various orders of derivatives in the effective action.
For example, $R^2$ terms come in at order $1/k^2$:
\ba
\int {d^4 x \over V_{\rm \, fl.\, sp.}}\,\sqrt{G} \,
R^{\mu\nu\rho\sigma}\, R_{\mu\nu\rho\sigma}&=&
{1\over k^2}{\sqrt{1-2H^2}\over 3}
\left(\l+{1\over \l}\right)\times \cr
&&\times \bigg(
\left(33+44H^2+132H^4\right)
\left(\l^2+{1\over \l^2}\right)16\left(3+4H^2\right)
\bigg)\cr
&=&
{1\over k^2}\left(12 +{\f^2\over 2} +{47 \rr^2\over 2}+\cdots\right)
\, .
\label{31}
\ea
Putting together eqs. (\ref{5}) and (\ref{25})--(\ref{28}), we obtain
the one-loop correction to the free energy in the effective field
theory:
\ba
\alpha'^2 F^{\rm \, effective}_{\rm \, one \ loop}&=&
{S_{\rm one \ loop} \over V_{\rm \, fl.\, sp.}}
\cr
&=&{1\over k}\bigg(2\,(3\zr+2\zf-\zh)
-(\zr+2\zf+\zh+4\za)\, {\f^2\over 4}
\cr
&&+(3\zr+2\zf-\zh)\, {\rr^2\over 4}+{\rm higher\ orders\ in
\ \f \ and \ \rr} \bigg)
+ {\cal O}\left({1\over k^2}\right)
 .
\label{30}
\ea

In order to determine the various renormalization constants, we
have to compare the effective field theory result (\ref{30}) with the
string calculation of the one-loop free energy given by eqs.
(\ref{32}) and (\ref{366}). We first note that the absence of a term
independent of $\f$ and $\rr$ in (\ref{366}) leads to
\be
3\zr+2\zf-\zh=0\, .
\label{37}
\ee
In turn, this relation implies through (\ref{30}) that it should not
be any $\rr^2$ term\footnote{Equation (\ref{31}) suggests, however,
that $\rr^2$ terms are present at the order $1/k^2$, which is again
in agreement with the string result (\ref{366}). This makes it
possible for the determination of the one-loop renormalization
constant $Z_{R^2}$, leading in particular to the gravitational
anomaly.} at order $1/k$ in (\ref{366}), which is indeed the
case.

Before proceeding further with the computation of the string-induced
renormalizations, it is interesting to observe that, independently of
any string-based consideration, relation (\ref{37}) is a consequence
of space-time supersymmetry. The argument is the following.
The tree plus one-loop action for the universal sector is given from
(\ref{1}) and (\ref{5}):
\ba
S_{\, {\rm tree \ \& \ one \ loop}}={1\over \alpha'}\int d^{4}x\,
\sqrt{G}\bigg(\!\!\!\!\!\!\!\!\!\!&&\left(e^{-2\Phi}+\zr\right)
R\cr
&&+4\left(e^{-2\Phi}+\zf\right)(\n\Phi)^2-
{1\over 12}\left(e^{-2\Phi}+\zh\right){H}^2\bigg)
\, .
\label{38}
\ea
By performing the transformation
\be
G_{\mu\nu}={1\over e^{-2\Phi}+\zr}\, g_{\mu\nu}\, ,
\ee
we can go to the Einstein frame, where the above action reads:
\ba
S^{\rm \  \ \ Einstein}_{\, {\rm tree \ \& \ one \ loop}}={1\over
\alpha'}\int d^{4}x\,
\sqrt{g}\Bigg(R\!\!\!\!\!\!\!\!\!\!&&-2\left(1-2\,
(\zf+2\zr)e^{2\Phi}\right)(\n\Phi)^2\cr
&&-
{e^{-4\Phi}\over
12}\left(1+(\zr+\zh)e^{2\Phi}\right){H}^2
\Bigg)
\, .
\label{39}
\ea
Only when relation (\ref{37}) is true does the action above, upon the
field redefinition
\be
\Phi'=\Phi-{\zf+2\zr\over 2}\, e^{2\Phi}+{\cal
O}\left(e^{4\Phi}\right)
\, ,
\ee
become the tree-level action in the Einstein frame, which is fixed
by supersymmetry.

In fact, this argument generalizes to higher orders in perturbation
theory, thus leading, at any order, to relations among
renormalization
constants similar to (\ref{37}). Let
\be
S={1\over \alpha'}\int d^{4}x\, \sqrt{G}\,
e^{-2\Phi}\left(F_R(\Phi)\, R+4F_{\phi}(\Phi)\, (\n\Phi)^2-
{1\over 12}\, F_{H}(\Phi)\, {H}^2\right)
\label{388}
\ee
be the all-order effective action for the universal sector in the
$\s$-model frame, where the functions $F_{R},F_{\phi}$ and $F_{H}$
have the perturbative expansion
\be
F_{K}^{\vphantom
)}(\Phi)=1+\sum_{n=1}^{\infty}Z_{K}^{(n)}e^{2n\Phi}\, .
\ee
Then $N=1$ supersymmetry implies that
\be
\log\big(F_{H}(\Phi)\, F_{R}(\Phi)\big)=-4\int_{-\infty}^{\Phi}dx
\left(\sqrt{3\left(1-{1\over 2}
{F_{R}'(x)\over F_{R}(x)}\right)^2-2\,
{F_{\phi}(x)\over F_{R}(x)}}-1\right)\, ,
\ee
which at the one-loop level leads precisely to (\ref{37}).

Let us now come back to the string computation and show that at one
loop
\be
\zr=\zh=0\, .
\label{44a}
\ee
In order to do this, we will go beyond the calculation that we
presented in section 2. Indeed, we have to study the two-point
amplitude
of graviton,
antisymmetric tensor and dilaton, at one loop. The piece quadratic in
momenta in such an amplitude determines the quadratic part of the
associated one-loop action, and therefore the corresponding
renormalization constant.
The relevant heterotic vertex operator is
\be
V(\epsilon,p)\propto \epsilon_{\mu\nu}\,
\big(\partial x^{\mu}+i(p\cdot \psi)\, \psi^{\mu}\big)\, \bar\partial
x^{\nu} e^{ip\cdot x}\, ,\label{41}
\ee
with $p^2=0$.
For comparison, the vertex operator of a gauge boson is
\be
V^a_{i}(\epsilon,p)\propto \epsilon_{\mu}^{\vphantom i}\big(\partial
x^{\mu}_{\vphantom i}
+i(p\cdot \psi)\, \psi^{\mu}_{\vphantom i}\big)\, \bJ^{a}_i \,
e^{ip\cdot x}
\, .\label{42}
\ee
The two-point $S$-matrix element on the torus is (up to an overall
normalization)
\be
S_{1\to 2}\propto \int_{\cal F}{d^2\t\over (\im )^2}
\int d^2 z \left\langle
V\big(\epsilon^{(1)},p^{(1)}\big)(z,\bar z)\;
V\big(\epsilon^{(2)},p^{(2)}\big)(0)
\right\rangle \, ,
\label{43}\ee
with
$p^{(1)\, \mu}_{\vphantom m}\, p^{(1)}_{\, \mu}=
 p^{(2)\, \mu}_{\vphantom m}\, p^{(2)}_{\, \mu}=
{\bf p}^{(1)}_{\vphantom m}+{\bf p}^{(2)}_{\vphantom m}=0$.
It is known that such an on-shell amplitude is zero, which is
consistent with the fact that there is no one-loop mass shift for
such massless particles. However, for our purposes, we have to go off
shell in order to pick out the terms quadratic in momenta. There
is such a prescription \cite{minahan,stie}, which amounts to keeping
$p^{(1)\, \mu}_{\vphantom m}\, p^{(1)}_{\, \mu}=
 p^{(2)\, \mu}_{\vphantom m}\, p^{(2)}_{\, \mu}=0$
but allowing
${\bf p}^{(1)}_{\vphantom m}+{\bf p}^{(2)}_{\vphantom m}$ to be
arbitrary, without destroying conformal or modular invariance.
Furthermore,
due to $N=1$ supersymmetry, the only term that contributes a non-zero
result
in (\ref{43}) is the one containing the four worldsheet fermions;
since we are interested in terms with two derivatives, we can set the
exponential $e^{ip\cdot x}$ to 1.
Thus
\ba
S_{1\to 2}&\propto
&
 \int_{\cal F}{d^2\t\over (\im )^2}
\int d^2 z\,
\epsilon^{(1)}_{\eta\kappa}\,
\epsilon^{(2)}_{\mu\nu}\,
p^{(1)}_{\, \rho}\,
p^{(2)}_{\, \sigma}\,
\big\langle
\psi^{\rho}(z)\,
\psi^{\eta}(z)\,
\psi^{\s}(0)\,
\psi^{\mu}(0)
\big\rangle
\left\langle
\bar \partial x^{\kappa}(\bar z) \,
\bar \partial x^{\nu}(0)
\right\rangle
\cr
&&+{\cal O}\left(p^4\right)\ \ \ \ \ \ \ \ \ \ \ \ \ \
\ \ \ \ \ \ \ \ \ \ \ \ \ \ \ \ \ \ \ \ \ \ \
\ \ \ \ \ \ \ \ \ \ \ \ \ \  \ \ \ \ \ \ \ \ \ \ \ \ \ \ \ \
{\rm(off\ shell)}\, .
\label{44}
\ea
It is obvious from the above expression, that the integrand is a
total
holomorphic derivative of a function that is periodic and regular on
the torus.
This implies that this integral vanishes, which proves relation
(\ref{44a}).
We should note here that, for the gravitational sector, the relevant
integrals
are finite in the infra-red  so that no regularization is needed.

Another way to see the vanishing is to compare with the gauge-field
case, where
the associated amplitude has the form
\ba
S_{i,a \,  \to \, j,b}&\propto
&
\int_{\cal F}{d^2\t\over (\im )^2}
\int d^2 z\,
\epsilon^{(1)}_{\, \mu}\,
\epsilon^{(2)}_{\, \nu}\,
p^{(1)}_{\, \rho}\,
p^{(2)}_{\, \sigma}\,
\big\langle
\psi^{\rho}_{\vphantom j}(z)\,
\psi^{\mu}_{\vphantom j}(z)\,
\psi^{\s}_{\vphantom j}(0)\,
\psi^{\nu}_{\vphantom j}(0)
\big\rangle
\left\langle
\bJ^{a}_i(\bar z) \,
\bJ^{b}_j(0)
\right\rangle \cr
&&+{\cal O}\left(p^4\right)\ \ \ \ \ \ \ \ \ \ \ \ \ \
\ \ \ \ \ \
\ \ \ \ \ \ \ \ \ \ \ \ \ \ \ \ \ \ \ \ \ \ \ \ \ \ \ \
\ \ \ \ \ \ \ \ \ \ \ \ \ \ \ \,
{\rm(off\ shell)}\, .
\label{45}
\ea
The left-moving fermionic correlation function reduces to the
standard helicity trace while the integrated correlation function of
the right-moving affine currents gives ${\rm
Tr}\left(\bQ_i^2-{k_{i}\over 4\pi \im }\right)$, where $\bQ_i^2$ is
the gauge-group quadratic Casimir.
Upon inspection, we can see that there is a similar formula for the
gravitational case with $k_{i}=1$ and $\bQ_i$ replaced by the
(continuous)
momentum in a single direction.
For continuous momentum, ${\rm Tr}\left(\bP^2\right) \propto { 1\over
\im }$, and by modular invariance it must cancel the second term.

The above argument generalizes to higher loops in the
presence of $N=1$ supersymmetry. Again only the four-fermion term
contributes and the integrand is always a total derivative.
We can thus conclude that, to all orders in perturbation theory,
Newton's constant is not renormalized around heterotic ground states
with at least $N=1$ space-time supersymmetry. Similarly, there are no
perturbative corrections to the dilaton and antisymmetric tensor
kinetic terms.
\vskip 0.3cm
{\bf 4. One-loop gauge couplings and universal thresholds}

The one-loop correction to the gauge coupling can be calculated using
the results of the previous section. We will describe the general
structure of these corrections in the $\overline{DR}$ scheme for a
generic supersymmetric four-dimensional model, and will eventually
concentrate on a specific family
of $N=2$ ground states that are two-torus compactifications of
arbitrary $N=1$ theories in six dimensions. In that case, it turns
out
that the universal, i.e.
group-factor independent, part of the thresholds is {\it truly
universal:}
it does not depend on the $(4,0)$ internal conformal field theory
that is used to reach six dimensions starting from ten.

Equations
(\ref{30}), (\ref{37}) and (\ref{44a}) imply
\be
\alpha'^2 F^{\rm \, effective}_{\rm \, one \ loop}={1\over k}\left(
-\za\, \f^2
+{\rm higher\ orders\ in\ \f\ and \ \rr}\right)+
{\cal O}\left({1\over k^2}\right)\, ,
\label{46}
\ee
where $\za$ can be determined by
comparison with eqs. (\ref{32}) and (\ref{366}).
Note that the normalizations for the effective field theory are
chosen such that the highest roots of the group algebra have length
squared equal to 1 \footnote{These are the usual normalizations that
lead in particular to the tree-level relation $M_s={\gs \over
\sqrt{32\pi G_N}}$.}. Since our string computation was performed with
$\psi ^2 =2$, the net result for $\za$ reads:
\be
\za=
{i\over 16\pi^3 k_{i}}
\int_{\cal F}
{d^2\t\over \im}\, {\Gamma(k) \over|\eta|^4}
\sum_{a,b=0,1}
\partial_{\t}
\left({\vartheta{a\atopwithdelims[]b}\over\eta}\right)\,
\left(
\bP_i^2 - {k_{i}\over 4\pi{\im}}\right) C{a\atopwithdelims[]b}\, .
\label{47}
\ee
{}From (\ref{1}) and (\ref{5}) we can derive the effective one-loop
string-corrected coupling $g_{{\rm eff,\,} i}$:
\ba
{16\pi^2\over g_{{\rm eff,\,} i}^2}&=&k_{i}{16\pi^2\over \gs^2}+
16\pi^2 k_{i}\,\za  \cr
&=&k_{i}{16\pi^2\over \gs^2}+{i\over \pi}
\int_{\cal F}
{d^2\t\over \im}\, {\Gamma(k) \over|\eta|^4}
\sum_{a,b=0,1}
\partial_{\t}
\left({\vartheta{a\atopwithdelims[]b}\over\eta}\right)\,
\left(
\bP_i^2 - {k_{i}\over 4\pi{\im}}\right) C{a\atopwithdelims[]b}\,
.\label{48}
\ea
The latter has to be identified with the corresponding field theory
one-loop gauge coupling,  regulated
in the infra-red in a similar fashion as the string theory. As
expected,
the infra-red divergence cancels between string theory and field
theory results. The effective field theory has also to be supplied
with an
ultraviolet cut-off; expressing the field theory bare coupling in
terms of the running coupling $g_{i}(\mu)$, we obtain \cite{pr}:
\be
{16\pi^2\over g_{i}^2(\mu)} = k_{i}{16\pi^2\over \gs^2} +
 b_{i}\log {M_{s}^2\over \mu^2} +\Delta_{i}
\label{49}
\ee
in the $\overline{DR}$
scheme,
where
\be
b_{i}=
\lim_{\im\to\infty}{i\over \pi}\, {1 \over|\eta|^4}
\sum_{a,b=0,1}
\partial_{\t}
\left({\vartheta{a\atopwithdelims[]b}\over\eta}\right)\,
\left(
\bP_i^2 - {k_{i}\over 4\pi{\im}}\right) C{a\atopwithdelims[]b}
\label{50}
\ee
are the usual beta functions, and
\be
\Delta_{i}=\int_{\cal F}{d^2\t\over \im}
\left(
{i\over \pi}
\, {1 \over|\eta|^4}
\sum_{a,b=0,1}
\partial_{\t}
\left({\vartheta{a\atopwithdelims[]b}\over\eta}\right)\,
\left(
\bP_i^2 - {k_{i}\over 4\pi{\im}}\right) C{a\atopwithdelims[]b}-
b_{i}
\right)
+b_{i}\log {2\,e^{1-\gamma}\over \pi\sqrt{27}}
\, .
\label{51}
\ee
Part of the threshold correction is universal
(gauge-group independent). We can thus split (\ref{51}) as
\be
\Delta_{i}=\hat \Delta_{i}-k_{i}\,Y \, .
 \label{52}
\ee
The universal piece $Y$ contains, among other things, contributions
from
the universal sector (gravity in particular).
Such contributions are absent in grand unified theories.
Thus $Y$ is a finite correction to the ``bare" string coupling $\gs$.
Moreover it is infra-red-finite, which in particular means that it
remains
finite when extra states become massless at some special values of
the moduli.
Thus we can write (\ref{49}) as
\be
{16\pi^2\over g_{i}^2(\mu)}=k_{i}{16\pi^2\over g_{\rm renorm}^2}+
b_{i}\log {M_{s}^2\over \mu^2} +\hat \Delta_{i}\, ,
\label{53}
\ee
where we have defined a ``renormalized" string coupling by
\be
g_{\rm renorm}^2={\gs^2\over 1-{Y\over 16\pi^2}\,\gs^2 }\, .
\label{55}
\ee
Moreover, for $N=2$ ground states
\be
\hat\Delta_{i}=b_{i}\, \Delta\, ,
\ee
where $\Delta$ does not depend on the group factor.
Then (\ref{53}) becomes
\be
{16\pi^2\over g_{i}^2(\mu)}=
k_{i}{16\pi^2\over g_{\rm renorm}^2}+b_{i}\log {M_{s}^2\,
e^{\Delta}\over \mu^2}
\, ,
\label{54}
\ee
and the couplings are unified at $\mu=M_{s}\,e^{\Delta\over2}$.

Let us now concentrate on a particular class of $N=2$ ground states,
namely those that come from toroidal compactification of generic
six-dimensional
$N=1$ string  theories\footnote{Note that this is not the most
general four-dimensional $N=2$ theory.}. Here, there is a universal
two-torus, which provides the (perturbative) central charges of the
$N=2$ algebra.
Therefore (\ref{51}) becomes
\be
\Delta^{N=2}_{i}=\int_{\cal F}{d^2\t\over \im}
\Bigg({\Gamma_{2,2}\left(T,U,\bT,\bU\right)\over \bar \eta^{24}}
\left(\bP_{i}^2-{k_{i}\over 4\pi\im}
\right)\overline{\Omega}-b_{i}
\Bigg)+b_{i}\log {2\, e^{1-\gamma}\over \pi\sqrt{27}}\, ,
\label{56}
\ee
where $T$ and $U$ are the complex moduli of the two-torus,
$\overline{\Omega}$ is an antiholomorphic function
and
\ba
\Gamma_{2,2}\left(T,U,\bT,\bU\right)=
\sum_{{\mbox{\footnotesize\bf m}},{\mbox{\footnotesize\bf n}}\in Z}
\exp
\bigg(\!\!\!\!\!\!\!\!\!\!&&
2\pi i\t\left(
m_{1}\, n^{1}+m_{2}\, n^{2}
\right)\cr &&-
{\pi\im\over \Im T \Im U}
\left|T n^{1} + TUn^{2}-Um_1+m_2\right|^2
\bigg)\, .
\label{60}
\ea
{}From (\ref{56}), we observe that the function
\be
\bF_{i}={1\over \bar \eta^{24}}
\left(\bP_{i}^2-{k_{i}\over 4\pi\im}
\right)
\overline{\Omega}
\label{57}
\ee
is modular invariant.
Consider the associated function that appears in the
$R^2$-term renormalization (see eq. (\ref{366}) or ref. \cite{antg}
for more details),
\be
\bF_{\rm grav}=
{\widehat{E}_2\over 12}\, {\bOmega \over \bar \eta^{24}}
={1\over \bar \eta^{24}}\left({i\over
\pi}\partial_{\tb}\log\bar\eta-{1\over 4\pi\im}\right)\bOmega
\, ,
\label{58}
\ee
which is also modular invariant, and eventually leads to the
gravitational anomaly.
The difference $\bF_{i}-k_{i}\, \bF_{\rm grav}$ is an antiholomorphic
function, which is modular invariant. It has at most a simple pole at
$\t\to i\infty$ (associated with the heterotic unphysical tachyon)
and is finite inside the moduli space of the torus. This implies that
\be
\bF_{i}=k_{i}\, \bF_{\rm grav}+A_i\, \bj(\tb)+B_i\, ,
\label{59}
\ee
where $A_i$ and $B_i$ are constants to be determined, and
$j(\t)={1\over q}+744+{\cal O}(q)$,
$q=\exp(2\pi i\t)$ is the standard $j$-function.
The modular invariance of $\bF_{\rm grav}$ implies that $\bOmega$ is
a modular form of weight 10, which is finite inside the moduli
space.
This property fixes
\be
\bOmega =\xi \,  \bE_{4}\, \bE_{6}\, ,
\label{61}
\ee
where $E_{2n}$ is the $n$th Eisenstein series:
\be
E_{2}=
{12\over i \pi}\partial_{\t}\log \eta
=1-24\sum_{n=1}^{\infty}{n\, q^n\over 1-q^n}
\, ,
\label{61b}
\ee
\be
E_{4}=
{1 \over 2}\left(
{\vartheta}_2^8+
{\vartheta}_3^8+
{\vartheta}_4^8
\right)
=1+240\sum_{n=1}^{\infty}{n^3q^n\over 1-q^n}
\, ,
\ee
\be
E_{6}=
\frac{1}{2}
\left({\vartheta}_2^4 + {\vartheta}_3^4\right)
\left({\vartheta}_3^4 + {\vartheta}_4^4\right)
\left({\vartheta}_4^4 - {\vartheta}_2^4\right)
=1-504\sum_{n=1}^{\infty}{n^5q^n\over 1-q^n}
\, .
\ee
Putting everything together in (\ref{56}) we obtain:
\be
\Delta^{N=2}_{i}=\int_{\cal F}{d^2\t\over \im}\,
\Bigg(\Gamma_{2,2}\left(T,U,\bT,\bU\right)
\left({\xi k_{i}\over 12}{\widehat{E}_{2}\,
\bE_{4}\, \bE_{6}\over \bar \eta^{24}}+A_i \, \bj+B_i\right)-b_{i}
\Bigg)+b_{i}\log {2\, e^{1-\gamma}\over \pi\sqrt{27}}
\, .
\label{62}
\ee

There are two constraints that allow us to fix the constants $A_i,
B_i$.
The  first is that the $1/q$ pole is absent from the group trace,
which gives
\be
A_i=-{\xi k_{i}\over 12}\, .
\label{64}\ee
The second is (\ref{50}), which implies
\be
744\, A_i+B_i-b_{i}+ k_i\,  b_{\rm grav} = 0\, ,
\label{63}\ee
where
\be
b_{\rm grav} =  \lim_{\im \to\infty}\bF_{\rm grav} = -22\,\xi
\label{bla}
\ee
is the gravitational anomaly in units where  a hypermultiplet
contributes
${1\over 12}$ \cite{antg}.
Plugging (\ref{64})--(\ref{bla}) in (\ref{62}), we finally obtain:
\ba
\Delta^{N=2}_{i}&=&
b_{i}\left(
\log {2\, e^{1-\gamma}\over \pi\sqrt{27}}+\int_{\cal F}{d^2\t\over
\im}\, \Big(\Gamma_{2,2}\left(T,U,\bT,\bU\right)-1\Big)
\right)\cr
&&+{\xi k_{i}\over 12}\int_{\cal F}{d^2\t\over \im}\,
\Gamma_{2,2}\left(T,U,\bT,\bU\right)
\left({\widehat{E}_{2}\, \bE_{4}\, \bE_{6}\over \bar
\eta^{24}}-\bj+1008\right)
\, .
\label{65}
\ea
The first integral in (\ref{65}) was computed explicitly in
\cite{dkl} and recently generalized in \cite{hm}:
\be
\int_{\cal F}{d^2\t\over \im}
\Big(\Gamma_{2,2}\left(T,U,\bT,\bU\right)-1\Big)=
-\log\left(\big|\eta(T)\big|^4 \big|\eta(U)\big|^4
\Im T  \Im U\right)-\log{8\pi \, e^{1-\gamma}\over \sqrt{27}}\, .
\label{66}
\ee
Therefore, as advertised above, we can write
\be
\Delta^{N=2}_{i}=b_{i}^{\vphantom N}\,
\Delta-k_{i}^{\vphantom N} \, Y\, ,
\ee
with
\be
\Delta=-\log\left(4\pi^2 \big|\eta(T)\big|^4 \big|\eta(U)\big|^4
\Im T  \Im U\right)
\label{67}
\ee
and
\be
Y=-{\xi \over 12}\int_{\cal F}{d^2\t\over \im}\,
\Gamma_{2,2}\left(T,U,\bT,\bU\right)\, \Bigg(
\left(\bE_{2}-{3\over \pi\im}\right){\bE_{4}\, \bE_{6}\over \bar
\eta^{24}}-\bj+1008
\Bigg)
\label{68}
\ee
(we have used eq. (\ref{377})).
This form of the universal term was determined for the case of
$Z_{2}\times Z_{2}$ orbifolds in \cite{pr} and further generalized
for a larger class of models in \cite{kkpr}.

The coefficient $\xi$ can be related to the number of massless
vector multiplets $N_{V}$ and hypermultiplets $N_{H}$ via the
gravitational
anomaly ($b_{\rm grav}$), which
can also be computed from the low-energy theory of massless states.
In units where a scalar contributes $1$,
the graviton contributes $212$, the antisymmetric tensor $91$, the
gravitino $-{233\over 4}$, a vector~$-13$ and a Majorana fermion
$7\over 4$; therefore
the $N=2$ supergravity multiplet contributes $212-2{233\over
4}-13={165\over 2}$, the tensor multiplet contributes $-13+2{7\over
4}+1+91={165\over 2}$,
a vector multiplet $-13+2{7\over 4}+2=-{15\over 2}$ and a
hypermultiplet
$2{7\over 4}+4={15\over 2}$.
Thus in the units of (\ref{bla}),
\be
b_{\rm grav} = {22 - N_V +N_H\over 12}\, ,
\ee
and hence
\be
\xi=-{1\over 264}\left(22-N_{V}+N_{H}\right)\, .
\label{70}
\ee
Therefore, the universal contribution (\ref{68}) reads:
\be
Y={22-N_V+N_H\over 3168}\int_{\cal F}{d^2\t\over \im}\,
\Gamma_{2,2}\left(T,U,\bT,\bU\right)\,
\Bigg(\left(\bE_2-{3\over \pi\im}\right){\bE_4\, \bE_6\over \bar
\eta^
{24}}-\bj+1008
\Bigg)\, .
\label{71}
\ee

As an example, let us consider
the case of the $Z_{2}$ orbifold, where we have a gauge group
$E_{8}\times
E_7\times SU(2)\times U(1)^2$ and thus $N_V=386$.
The number of massless hypermultiplets is $N_H=628$.
Using these numbers in (\ref{70}) we obtain in this case $\xi=-1$. As
expected by supersymmetry, the corresponding universal threshold is
twice as big as a single-plane contribution of the symmetric
$Z_{2}\times Z_{2}$ orbifold.

We come finally to the
important observation that the result $\xi=-1$
applies to more general situations than the above example. One
can indeed show that $N_{H}-N_{V}$ is a {\it universal constant}
for the whole class of four-dimensional $N=2$ models obtained
by toroidal compactification of {\it any} $N=1$ ground state
in six dimensions. The argument is the following. From the
six-dimensional point of view,
the models at hand must obey an anomaly-cancellation
constraint,  which reads:~$\left. N_{H}-N_{V}\right|_{\rm six \
dim}=244$,
and does not depend on the kind of compactification that has been
performed from ten to six
dimensions\footnote{Actually, this constraint, which ensures that
$\Tr R^4$ vanishes, holds even when there occurs a symmetry
enhancement originated from non-perturbative effects, provided the
number of tensor multiplets remains $N_T = 1$. Note that this
six-dimensional anomaly-cancellation constraint is also used
in \cite{uu}, in relation to four-dimensional quantities.}
\cite{sch}.
After two-torus compactification to four dimensions,
two extra $U(1)$'s appear, leading to the relation
\be
N_{H}-N_{V}=242
\label{anom}
\ee
between the
numbers of vector multiplets and hypermultiplets.
In turn, eq. (\ref{70}) implies that for this class of
ground states $\xi=-1$, as advertised previously.
As a consequence,
all $N=2$ models under consideration have equal
universal thresholds, given by (\ref{71})
and (\ref{anom}).
\vskip 0.3cm
{\bf 5. Conclusions}

We have applied the background field method to analyze the response
of a string, supplied with an appropriate curvature-induced infra-red
cut-off, to magnetic and gravitational marginal deformations. This
has allowed us to obtain the exact (i.e. to all orders in $\ap$)
genus-1 free energy of the string as an expansion with respect to
the space-time curvature parameter $1/k$. Comparison with the
effective field theory $\s$-model action has led to definite results
for the one-loop corrections to gauge and gravitational couplings.
Indeed, we have demonstrated the absence  of corrections to Newton's
constant for supersymmetric ground states, and argued how this result
can be extended to all orders in $\gs$. We have then derived the full
one-loop gauge coupling in the $\overline{DR}$ scheme, for $N=1$
supersymmetric theories.

For the class of $N=2$ four-dimensional theories that come from
torus compactification of six-dimensional $N=1$ ground states, using
the relation between gauge and $R^2$-term renormalizations, we have
obtained an explicit formula for the universal part of the threshold
corrections. It is quite remarkable that the latter turns out to be
related to the quantity $N_{H}-N_{V}$, which is {\it fully
determined}
as a consequence of {\it the anomaly cancellation} (gauge,
gravitational
and mixed) in the underlying six-dimensional theory. Therefore,
the whole class of models under consideration have
equal universal thresholds. Note, however, that although
$N_{H}-N_{V}$ is not expected
to receive any non-perturbative contribution as long as $N_T = 1$,
the universal thresholds in general are.

We would like, finally, to note that the above results can be further
generalized to the contributions of the $N=2$ supersymmetric sectors
in $N=1$ string models \cite{kkpr}. In these models, which are
phenomenologically interesting, the universal thresholds will play a
role for the issue of string unification \cite{pr}.
\vskip 0.3cm
\centerline{\bf Acknowledgements}

C. Kounnas was  supported in part by EEC contracts
SC1$^*$-0394C and SC1$^*$-CT92-0789. P.M. Petropoulos
acknowledges financial
support from the EEC contracts CHRX-CT93-0340  and SC1-CT92-0792.
J. Rizos would like to thank the CERN Theory Division  for
hospitality
and  acknowledges financial
support from the EEC contract {ERBCHBGCT940634}.

\end{document}